\documentclass[12pt,a4]{article}

\newcommand{\diff}{\mathrm d} 
\newcommand{\ci}{\mathrm i} 
\newcommand{\ce}{\mathrm e} 
\usepackage{graphicx,color}

\title{Eigenvalue Problem  
of Scalar Fields in BTZ Black Hole Spacetime 
\footnote{The part of this work was presented in   
The 17th Workshop on General Relativity and Gravitation 
\cite{jgrg17}. }}
\author{Maiko KUWATA \thanks{kuwata@asuka.phys.nara-wu.ac.jp},  
Masakatsu KENMOKU \thanks{kenmoku@asuka.phys.nara-wu.ac.jp} \\
Department of Physics, Nara Women's University, Nara 630-8506, Japan
\\and 
Kazuyasu SHIGEMOTO \\
Tezukayama University, Nara 631-8501, Japan
\thanks{shigemot@tezukayama-u.ac.jp}}

\date{\empty}
\begin{document} 
\maketitle

%-------------------------------------------
\abstract{
We studied the eigenvalue problem of scalar fields in the
(2+1)-dimensional BTZ black hole spacetime. 
The Dirichlet boundary condition at infinity and the 
Dirichlet or the Neumann boundary condition at the horizon are imposed.   
Eigenvalues for normal modes are characterized by the principal 
quantum number $(0<n)$ and the azimuthal quantum number 
 $(-\infty<m< \infty)$. 
Effects to eigenvalues  
of the black hole rotation and of the scalar field mass are 
studied explicitly. 
Relation of the black hole rotation to the super-radiant instability 
is discussed.   
}

%-------------------------------------------
\section{Introduction}
%-------------------------------------------
Black hole physics is making progress theoretically and observationally. 
Especially interactions among black holes and matter fields 
are important in the scattering problem, in the quasi-normal modes, 
in the black hole thermodynamics and in others \cite{birrell001}. 
As to black hole thermodynamics, 
black holes are considered as thermal objects 
with the special temperature and the entropy 
\cite{bekenstein001,bardeen001,hawking001}. 
The microscopic derivation of black hole thermodynamics  
is desirable to make clear the black hole dynamics. 
Many attempts from the string theory, from the conformal field theory, 
from the brick wall model and from others have been done. 
Among them, the brick wall model by 't Hooft is  
the standard method of 
statistical mechanics to derive the black hole thermodynamics 
\cite{thooft001}.   
The attempts for non-rotating black holes are successful 
but the attempts for rotating black holes are problematic 
because of the super-radiance instability 
\cite{bardeen002,cardoso001,mukohyama001}. 
In order to solve the super-radiance problems clearly, 
the exact analysis is required.  

The (2+1)-dimensional anti-de Sitter spacetime is important, 
because it has exact rotating black hole solution 
by Banados, Teitelboim and Zaenlli \cite{btz001}.
The qusinormal modes in the BTZ black hole spacetime are found analytically 
\cite{birmingham001,cardosolemos002}. 
The conformal filed theory approach to the BTZ black hole model 
has been done in the classical mechanics and in the quantum mechanics 
\cite{carlip001,ksgupta001,witten001,rkgupta001}.  
The brick wall model in the BTZ spacetime was studied extensively 
but there were the problem of  divergence in taking the statistical sum due to 
the super-radiance instability 
\cite{ichinose001,swkim001,fatibene001,ho001}.

As a related topics, 
normal modes for the anti-de Sitter spacetime were studied 
in the framework of the gravitational perturbation \cite{birmingham002}
by using the gauge invariant formalism 
\cite{ishibashi001,kodama001} in any dimension 
but the effect of the black hole rotation was not considered.  

The purpose of this paper is 
to construct quantum states of matter fields explicitly 
in the rotating black hole spacetime, 
which will give the basic microscopic states for the 
black hole thermodynamics. 
Our model is the scalar field model in the (2+1)-dimensional BTZ 
black hole spacetime, which will provide the  
exact analysis to make clear the effect of the  black hole rotation 
to eigenvalues of normal modes.    
We impose the Dirichlet boundary condition at infinity 
and the Dirichlet or the Neumann boundary condition at the horizon 
to get eigenvalues and eigenfunctions explicitly.

This paper plays the complementary role to 
our previous work on the general analysis about normal 
mode \cite{kenmoku001}. 
 
The organization of this paper is the following. 
In section 2, notations and definitions of the BTZ black hole 
and the equation of scalar fields in this spacetime are explained. 
In section 3, the boundary conditions are imposed 
and eigenvalue equations for scalar fields are derived. 
In section 4, eigenvalue equations are solved numerically and 
obtain eigenvalues and eigenfunctions explicitly.    
Results are summarized in the final section.

%-------------------------------------------
\section{Scalar fields in the BTZ spacetime}
%-------------------------------------------
In this section, we prepare definitions and notations for 
following main sections. We take the natural unit  $\hbar=c=1$ 
and the  gravitational constant $G=1/8$ throughout this paper.

For the negative cosmological constant
$(\Lambda=-1/\ell^2)$ 
in the (2+1)-dimension, the exact black hole metric is obtained by  
{{Banados, Teitelboim and Zanelli (BTZ)}} \cite{btz001}: 
\begin{eqnarray}
ds^2&=&g_{tt}dt^2+g_{\phi \phi}d\phi^2+2g_{t\phi}dtd\phi+g_{rr}dr^2\ ,
\nonumber\\
g_{tt}&=&M-\frac{r^2}{\ell^2}\ , \ \   g_{t\phi}=-\frac{J}{2}\ , \ \  
g_{\phi\phi}=r^2\ , 
\ \  g_{rr}=\left({-M+\frac{J^2}{4r^2}+\frac{r^2}{\ell^2}}\right)^{-1}
\nonumber \ ,
\end{eqnarray}
where $M$ and $J$ are the mass and the angular momentum 
of the black hole respectively. 
Outer and inner horizon are defined by:  
\begin{eqnarray}
r_{\pm}^2=\frac{M\ell^2}{2}\left({1\pm\sqrt{1-\frac{J^2}{M^2\ell^2}}}\right)
 \ .
\end{eqnarray}
The event horizon is outer horizon $r_{+}$.
The action of the complex scalar field $\Phi(x)$ with mass $\mu$ is 
\begin{eqnarray}
I_{\rm scalar}=-\int {dt dr d\phi} \sqrt{-g}
\left(g^{\mu\nu} \partial_{\mu}\Phi^* (x)\partial_{\nu}\Phi(x)
+\frac{\mu}{\ell^2}\Phi^*(x)\Phi(x)\right) \ .
\end{eqnarray}
The scalar field is written in the form of separation of variables 
 $\Phi = \ce^{- \ci \omega t + \ci m \phi} R(r)$ 
with the frequency $\omega$ and the azimuthal angular momentum $m$. 
Then the equation for the radial wave function $R(z)$ is obtained : 
\begin{eqnarray}
\left( g_{rr}(\omega-\frac{J}{2r^2}m)^2-\frac{m^2}{r^2}
+\frac{1}{r}\partial_{r}\frac{r}{g_{rr}}
\partial_{r}-\frac{\mu}{\ell^2} \right)R(r)=0 \ , 
\end{eqnarray}
with the boundary condition:
\begin{eqnarray}\label{boundary:001}
\sqrt{-g}g^{rr}\left(\delta R(r)^{*} \partial_{r}{R(r)}
+\partial_{r} R(r)^{*}\delta{R(r)}\right)\mid_{\rm boundary}=0
\ ,
\end{eqnarray}
where $\delta R $ is the variation of $R$.
Introducing the new radial variable $z$ and the new radial function $F(z)$ as
\begin{eqnarray}
z = \frac{r^2-r_+^2}{r^2-r_-^2}\ ,\ \  \ F(z) = z^{i\alpha}(1-z)^{-\beta}
 R(z)\ ,  
\end{eqnarray}
the hypergeometric differential equation is obtained : 
\begin{eqnarray}
	z(1-z)\frac{\diff^2 F}{\diff z^2}
	+ \left(c - (1+a+b)z \right)
          \frac{\diff F}{\diff z}
	- ab F = 0 \ .
\end{eqnarray}
The parameters $a$, $b$, $c$ are defined :
\begin{eqnarray}
a=\beta-i\frac{\ell^2}{2(r_{+}+r_{-})}\left(\omega+\frac{m}{\ell}\right) \ , \ 
b=\beta-i\frac{\ell^2}{2(r_{+}-r_{-})}\left(\omega-\frac{m}{\ell}\right) \ , \
c=1-2i\alpha \ ,
\label{parameter:001}
\end{eqnarray}
\begin{eqnarray}
\alpha = \frac{\ell^2r_{+}}{2(r_{+}^2-r_{-}^2)}
(\omega -\Omega_{\rm H}m) \ , \ \
\beta = \frac{1 - \sqrt{1 + \mu}}{2}\ ,  
\end{eqnarray}
where $\Omega_{\rm H}=J/2r_{+}^2$ is the angular velocity at the horizon.  
The general solution of the hypergeometric differential equation 
is expressed by a linear combination of two independent solutions   
at the horizon or at infinity. 

%------------------------------
\section{The eigenvalue problem of the scalar field} 
%------------------------------
In this section, we set up the boundary conditions 
at infinity and at the horizon 
in order to satisfy Eq.(\ref{boundary:001})
to obtain eigenvalues and eigenfunctions for the scalar fields. 

First we impose the Dirichlet boundary condition at infinity  
because BTZ solution is asymptotic AdS spacetime: 
\begin{eqnarray}
R_{\infty}
=\frac{z^{-i\alpha}(1-z)^{\beta}(1-z)^{c-a-b}}
{\Gamma(c-a-b+1)}F(c-a,c-b,c-a-b+1;1-z) \ .
\label{radial:001}
\end{eqnarray}
Near horizon, this solution is also expressed as 
incoming and outgoing waves to the black hole as: 
\begin{eqnarray}
R_{\infty}
&=&
\frac{\Gamma(1-c)}{\Gamma(1-a)\Gamma(1-b)}R_{r_{+},{\rm in}}
+ 
\frac{\Gamma(c-1)}{\Gamma(c-a)\Gamma(c-b)}R_{r_{+},{\rm out}}\ ,  
\end{eqnarray}
where the ingoing and outgoing waves are defined by the hypergeometric
function as   
\begin{eqnarray}
R_{r_{+},{\rm in}}&=&z^{-i\alpha}(1-z)^{\beta}F(a,b,c;z)\ , \nonumber \\ 
R_{r_{+},{\rm out}}&=&z^{i\alpha}(1-z)^{\beta}F(1+b-c,1+a-c,2-c;z)\ . 
\end{eqnarray}
Note that they are complex conjugate for each other 
for real value of $\omega$. 
Their approximate expression near the horizon is 
\begin{eqnarray}
R_{r_{+},{\rm in}}&\sim& 
\exp{\left(-i(\omega-\Omega_{\rm H}m)r_{*}-i\alpha_{0}(\omega)\right)}\
, 
\nonumber \\ 
R_{r_{+},{\rm out}}&\sim&
\exp{\left(i(\omega-\Omega_{\rm H}m)r_{*}+i\alpha_{0}(\omega)\right)}\ ,    
\end{eqnarray}
where the tortoise coordinate  
\begin{eqnarray}
r_{*}=\int^{r}dr g_{rr}\ =\frac{\ell^2}{2(r_+^2-r_-^2)}
\left(r_+\ln{\frac{r-r_+}{r+r_+}} -r_-\ln{\frac{r-r_-}{r+r_-}}\right),
\end{eqnarray}
and the phase function
\begin{eqnarray}
\alpha_{0}(\omega)=
\frac{\ell^2r_{+}(\omega-\Omega_{\rm H}m)}{2(r_{+}^2-r_{-}^2)}
\log{\frac{4r_{+}^2}{r_{+}^2-r_{-}^2}}\ ,
\end{eqnarray}
are introduced.

Next we impose the Dirichlet or the Neumann boundary condition  
at the horizon to obtain eigenvalue equations:  
\begin{itemize}
\item[(i)] The Dirichlet boundary condition at the horizon: \\
The radial wave function is required to satisfy 
\begin{eqnarray}
\left.\left[\frac{\Gamma(1-c)}
{\Gamma(1-a)\Gamma(1-b)}R_{r_{+},{\rm in}}
+\frac{\Gamma(c-1)}{\Gamma(c-a)\Gamma(c-b)}R_{r_{+},{\rm out}}
\right]\right|_{r_*=r_{*,{\rm H}}}=0\ , 
\end{eqnarray}
which leads the eigenvalue equation:
\begin{eqnarray}
(\omega-\Omega_{\rm H}m)r_{*, \rm H}
+\alpha_{0}(\omega)+\beta_{0}(\omega)
=-\pi \left(n+\frac{1}{2}\right)\ \ \ \mbox{\rm for} \ \ n=0,1,2,\cdots \ . 
\label{dbc:001}
\end{eqnarray}
\item[(ii)]The Neumann boundary condition at the horizon:\\
The radial wave function is required to satisfy 
\begin{eqnarray}
\label{siki:neumann}
\left.\left[\frac{\Gamma(1-c)}
{\Gamma(1-a)\Gamma(1-b)}R_{r_{+},{\rm in}}
-\frac{\Gamma(c-1)}{\Gamma(c-a)\Gamma(c-b)}R_{r_{+},{\rm out}}
\right]\right|_{r_*=r_{*,{\rm H}}}=0\ ,
\end{eqnarray}
which leads the eigenvalue equation:
\begin{eqnarray}
(\omega-\Omega_{\rm H}m)r_{*, \rm H}
+\alpha_{0}(\omega)+\beta_{0}(\omega)
=-\pi n\ \ \ \mbox{\rm for} \ \ n=0,1,2,\cdots \ .
\label{nbc:001}
\end{eqnarray}
\end{itemize}
In above equations, 
the phase function $\beta_{0}(\omega)$ is introduced:
\begin{eqnarray}
\beta_{0}(\omega)
=
\arg{\left(\frac{\Gamma(c-1)}{\Gamma(c-a)\Gamma(c-b)}\right)} \ , 
\end{eqnarray}
and the tortoise coordinate at the horizon $r_{*,{\rm H}}$ 
is expressed as 
\begin{eqnarray}
r_{*,\rm H}\simeq\frac{\ell^2}{2(r_+^2-r_-^2)}
\left(r_+\ln{\frac{\epsilon}{2r_+}} -r_-\ln{\frac{r_+-r_-}{r_++r_-}}\right),
\end{eqnarray}
with the small regularization parameter $\epsilon$. 
This regularization parameter plays the same role as that in the brick wall
model \cite{thooft001}. 
Each quantum state is specified 
by the principal quantum number $n$ and 
the azimuthal quantum number $m$.   
The eigenfunction $R_{\infty}$ in Eq.(\ref{radial:001}) 
is determined by the boundary condition 
and then all eigenvalues and eigenfunctions are determined.     

\section{The numerical result for the eigenvalue and the eigenfunction}
\begin{figure}[h]
\begin{center}
\includegraphics[height=4cm,width=8cm]{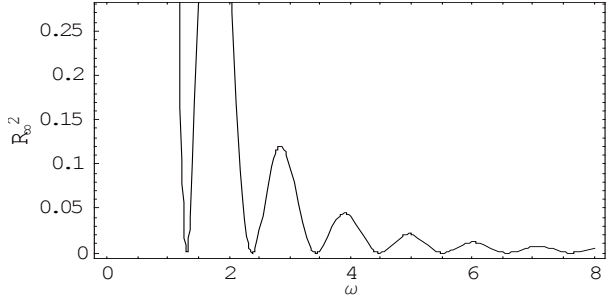}
\caption{Square of eigenfunction at the horizon with respect to $\omega$ }
\label{figure:eigenfunc}
\end{center}
\end{figure}

In this section, the numerical result for eigenvalues and 
eigenfunctions are shown. 
Here we show the 
square of absolute value of the eigenfunction
at the horizon with respect to the frequency $\omega$ in 
Fig.\ref{figure:eigenfunc}.  
In the numerical calculation, we set the parameter value 
for the black hole mass, angular momentum, the scalar mass and the 
cosmological parameter as 
\begin{eqnarray}
M=1, \  J=0, \  \mu=0, \ \ell=1 \ . 
\label{parameter:002} 
\end{eqnarray}
Throughout this numerical study, we set the regularization parameter as  
\begin{eqnarray}
z_{\epsilon}=0.01\ , \ \ {\mbox{\rm with}}\ \ z_{\epsilon}
=\frac{(r_{+}+\epsilon)^2-r_{+}^2}{(r_{+}+\epsilon)^2-r_{-}^2}
\simeq \frac{2r_{+}\epsilon}{r_{+}^2-r_{-}^2} \ .
\end{eqnarray}
In Fig.\ref{figure:eigenfunc}, 
the zeros of $R_{\infty}^2$ correspond to 
eigenvalues of the normal mode $\omega$ for the Dirichlet 
boundary condition at the horizon.  
Eigenvalues for the Neumann boundary condition at the horizon 
are between those for the Dirichlet boundary condition.

\subsection{The Dirichlet boundary condition at the horizon} 
We study the eigenvalue map with the eigenvalue points 
in $(m, \omega)$ plane 
for the Dirichlet boundary condition at the horizon.  

\subsubsection{The eigenvalue of no black hole rotation} 
\begin{figure}[h]
\begin{center}
\includegraphics[height=5cm,width=8cm]{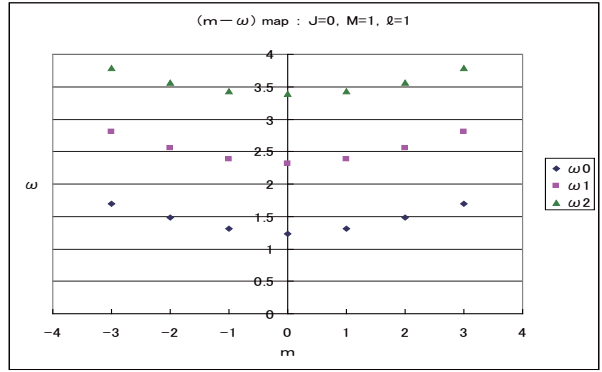} 
\caption{The eigenvalue map of no rotation for the 
Dirichlet B.C.}
\label{figure:dbc001}
\end{center}
\end{figure}

In the case of no black hole rotation $J=0$, 
the eigenvalue points $\omega_{n}$ for each fixed $n$ 
is shown in Fig.\ref{figure:dbc001}.    
Parameters are those of Eq.(\ref{parameter:002}).  
In the map, each eigenvalue point $(\omega_{n}, n=0,1,\dots)$ 
forms a convex curve with respect to the horizontal line.
In this $J=0$ case, the eigenvalue equation Eq.(\ref{dbc:001}) is invariant 
under the transformation $\omega \rightarrow \omega, m \rightarrow -m$
, which means that the eigenvalue $\omega_n $ is the even function 
of $m$. This is the origin of the convexity of the curve  
in $J=0$ case.

\subsubsection{The scalar mass effect}
\begin{figure}[h]
\centering
\includegraphics[height=5cm,width=8cm]{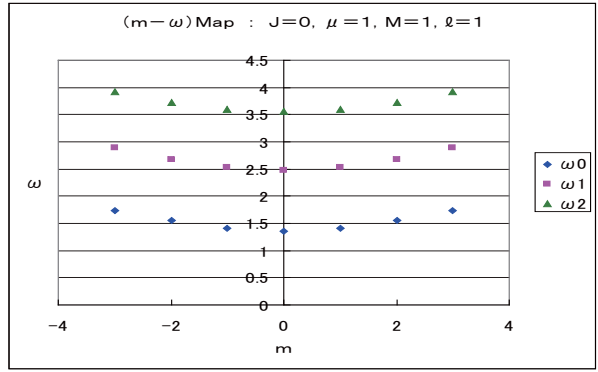}
\caption{ The scalar mass effect( $\mu=1$) of no rotation}   
\label{figure:dbc003}
\end{figure}

For the Dirichlet boundary condition, the scalar field mass
effect ($\mu\neq 0$) is studied 
in Fig.\ref{figure:dbc003} with parameter values 
\begin{eqnarray}
M=1, \ J=0, \ \ell=1, \  \mu=1 \ . 
\label{parameter:003}
\end{eqnarray}
We see from the map that the effect of the scalar mass term is 
to uniformly shift each eigenvalue to the larger value  
and the qualitative feature is similar to the massless case 
$(\mu=0)$.  

\subsubsection{The rotation effect of the black hole}
\begin{figure}[h]
\begin{center}
\includegraphics[height=5cm,width=8cm]{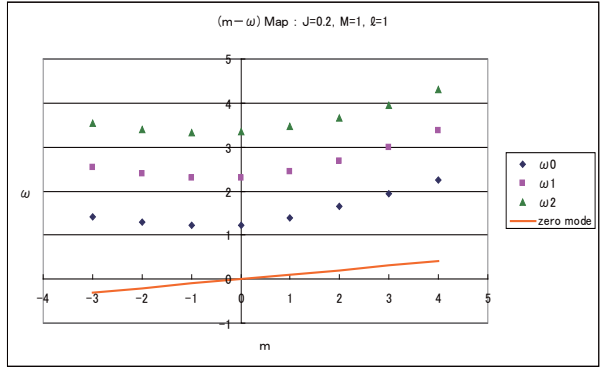} 
\caption{Rotation effect of black hole for Dirichlet B.C.}
\label{figure:dbc004}
\end{center}
\end{figure}

Next, for the Dirichlet boundary condition, we study 
the rotation effect of the black hole    
to the eigenvalue $(m, \omega)$ with parameters
\begin{eqnarray}
M=1, \  J=0.2, \  \mu=0,  \  \ell=1 \ . 
\label{parameter:003}
\end{eqnarray}
We can see in Fig. \ref{figure:dbc004} that points of 
eigenvalues $(m, \omega)$ rotate 
corresponding to the angler velocity $\Omega_{\rm H}$ 
in case of $J=0.2$ 
compared with the case of $J=0$ (See Fig. \ref{figure:dbc001}).
The zero mode line, which is defined:  
\begin{eqnarray}
\omega - \Omega_{\rm H}m=0 \ \ 
{\mbox {\rm with}}\ \ -\infty <m <\infty \ , 
\end{eqnarray}
is also shown in Fig.\ref{figure:dbc004}. We notice that 
all points of eigenvalues lie in the physical region 
$0<\omega - \Omega_{\rm H}m $. 

\subsection{The Neumann boundary condition at the horizon}
For the Neumann boundary condition at horizon, we study 
the eigenvalue map with eigenvalue points in $(m, \omega)$ plane.

\subsubsection{The eigenvalue of no black hole rotation}  
\begin{figure}[h]
\begin{center}
\includegraphics[height=5cm,width=8cm]{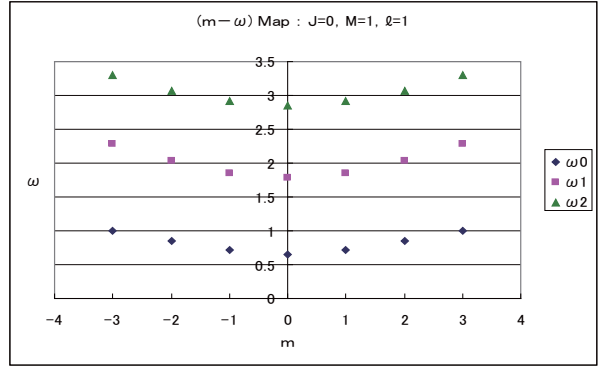} 
\caption{The eigenvalue map of no rotation for the Neumann B.C.}
\label{figure:nbc001}
\end{center}
\end{figure}

In case of no black hole rotation, 
the eigenvalue points $\omega_{n}$ for each fixed $n$ 
is shown in Fig.\ref{figure:nbc001} with 
same parameter values as those in Eq.(\ref{parameter:001}).  
For the Neumann boundary condition, each eigenvalue with fixed 
principal quantum number $n$ and azimuthal quantum number $m$ 
in Fig.{\ref{figure:nbc001}}
exist between the corresponding value of that for the Dirichlet 
boundary condition (See Fig.{\ref{figure:dbc001}}). 
We see that the ground state eigenvalues $\omega_{0}$ of $n=0$ 
for the Neumann  
boundary condition is lower than that for the
Dirichlet boundary condition because it has no node.  

\subsubsection{The rotation effect of the black hole}
\begin{figure}[h]
\begin{center}
\includegraphics[height=5cm,width=8cm]{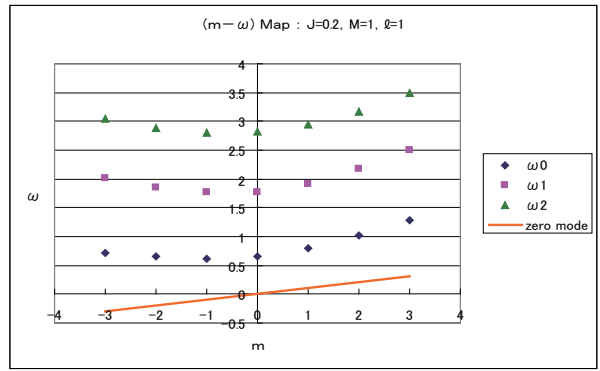} 
\caption{Rotation effect of black hole for the Neumann B.C.}
\label{figure:nbc002}
\end{center}
\end{figure}

For the Neumann boundary condition, we study the rotation effect 
of the black hole to points of eigenvalues $(m, \omega)$ with parameter 
values
\begin{eqnarray}
M=1,\  J=0.2, \ \mu=0, \  \ell=1 \ . 
\label{parameter:003}
\end{eqnarray}
We can see from the Fig.\ref{figure:nbc002} that points of  
eigenvalues $(m, \omega)$ rotate 
corresponding to the angler velocity $\Omega_{\rm H}$ 
in case of $J=0.2$ 
compared with no black hole rotation case in Fig.\ref{figure:nbc001} 
as in the case of the Dirichlet boundary condition.
The zero mode line separate all the eigenvalue region 
into two parts: one is the allowed physical region 
$0<\omega-\Omega_{\rm H}m$ with $-\infty <m <\infty$ 
and another is the unphysical region 
$0>\omega-\Omega_{\rm H}m$ with $-\infty <m <\infty$.   
  
%-------------------------------------------
\section{Summary and discussion}
%-------------------------------------------
We have studied the eigenvalue problem 
of scalar fields in the BTZ black hole spacetime. 
We imposed the Dirichlet boundary condition at infinity 
and the Dirichlet or the Neumann boundary condition at the horizon   
and found the explicit form of eigenfunctions 
and eigenvalues.
The main results are summarized in the following. 
\begin{itemize}
\item[(i)]
Eigenvalue equations in Eq.(\ref{dbc:001}) and Eq.(\ref{nbc:001}) are derived.
Eigenvalues for normal modes $\omega$ 
are characterized by the principal quantum number $n\ (n=0,1,2,\dots)$ 
and the azimuthal quantum number 
$m\ (-\infty, \dots , -1,0,1,2,\dots \infty)$.
The eigenfunction $R_{\infty}$ in Eq. (\ref{radial:001}) 
is determined with boundary condition and all eigenvalues 
and eigenfunctions are determined.  

\item[(ii)]  
The set of eigenvalues $(\omega, m)$ forms a convex curves 
in $(\omega, m)$ plane for fixed $n$.
For the Dirichlet boundary condition, we showed the convexity property 
of no black hole rotation, but this convexity property holds for other
cases too.

\item[(iii)]
The scalar mass effect is to uniformly shift each eigenvalue to 
the larger value but the qualitative feature is similar to 
the massless case $(\mu=0)$. 

\item[(iv)]
Points of eigenvalues $(\omega,m)$ rotate corresponding 
to the angular velocity 
$\Omega_{\rm H}$ for both the Dirichlet and the Neumann boundary conditions. 
Then the allowed physical eigenvalue region of $0< \omega$ for $J=0$ 
becomes $0< \omega -\Omega_{\rm H}m $ for $J > 0$. 
This result indicates that the super-radiant instability 
\cite{ichinose001,swkim001,fatibene001,ho001}
doesn't occur in the (2+1)-dimensional BTZ black hole spacetime. 
\end{itemize}

The explicit construction of eigenvalues and eigenfunctions 
studied in this paper plays the complimentary role with 
the general consideration of our previous paper 
\cite{kenmoku001}. 
Our method will be extend to the (3+1)-dimensional and more higher
dimensional cases \cite{kenmoku002}.

%-------------------------------------------

\end{document}